\documentclass[12pt]{iopart}

\usepackage{epsfig}
\usepackage{graphicx}
\begin{document}

\title[$\rho^0$ Production at High $p_T$ in Central
Au+Au and $p+p$ collisions at $\sqrt{s_{_{NN}}} = 200$ GeV in STAR]
{$\rho^0$ Production at High $p_T$ in Central
Au+Au and $p+p$ collisions at $\sqrt{s_{_{NN}}} = 200$ GeV in
STAR}

\author{P Fachini}

\address{ Brookhaven National Laboratory,
Upton, NY, 11973, USA}
\ead{pfachini@bnl.gov}
\begin{abstract}
The $\rho^0$ production at high-$p_T$ (5.0 $\leq p_T \leq$
10.0 GeV/$c$) measured in minimum bias $p+p$, Au+Au and central Au+Au
collisions in the STAR detector are
presented. The $\rho^0/\pi$ ratio measured in $p+p$ is
compared to PYTHIA calculations as a test of perturbative quantum
chromodynamics (pQCD) that describes reasonably well particle
production from hard processes. The $\rho^0$ nuclear modification
factor are also presented. In $p+p$ collisions, charged pions
and (anti-)protons are measured in the
range 5.0 $\leq p_T \leq$ 15.0 GeV/$c$ and the anti-particle to particle ratio and
the baryon to meson ratios of these hadrons are discussed.
\end{abstract}


\section{Introduction}

The study of inclusive hadron production at high $p_T$ in $p+p$
collisions provides information on perturbative QCD (pQCD), parton distribution
functions in the proton (PDF) and fragmentation functions (FF) of the partons.
In Au+Au collisions, high $p_T$ hadron production is a sensitive probe of the
strongly interacting QCD matter formed in these collisions. The transverse momentum spectra of $\pi$, $\rho^0$, and $p$ measured in
$p+p$ and central Au+Au collisions can be used to study the effect of energy
loss on fragmentation. These measurements can provide insight in the quantum
chromodynamics (QCD) predicted difference  between quark and gluon energy
loss.

\section{Analysis and Results}

\begin{figure}[htb]
\begin{minipage}[t]{85mm}
\includegraphics[height=14pc,width=18pc]{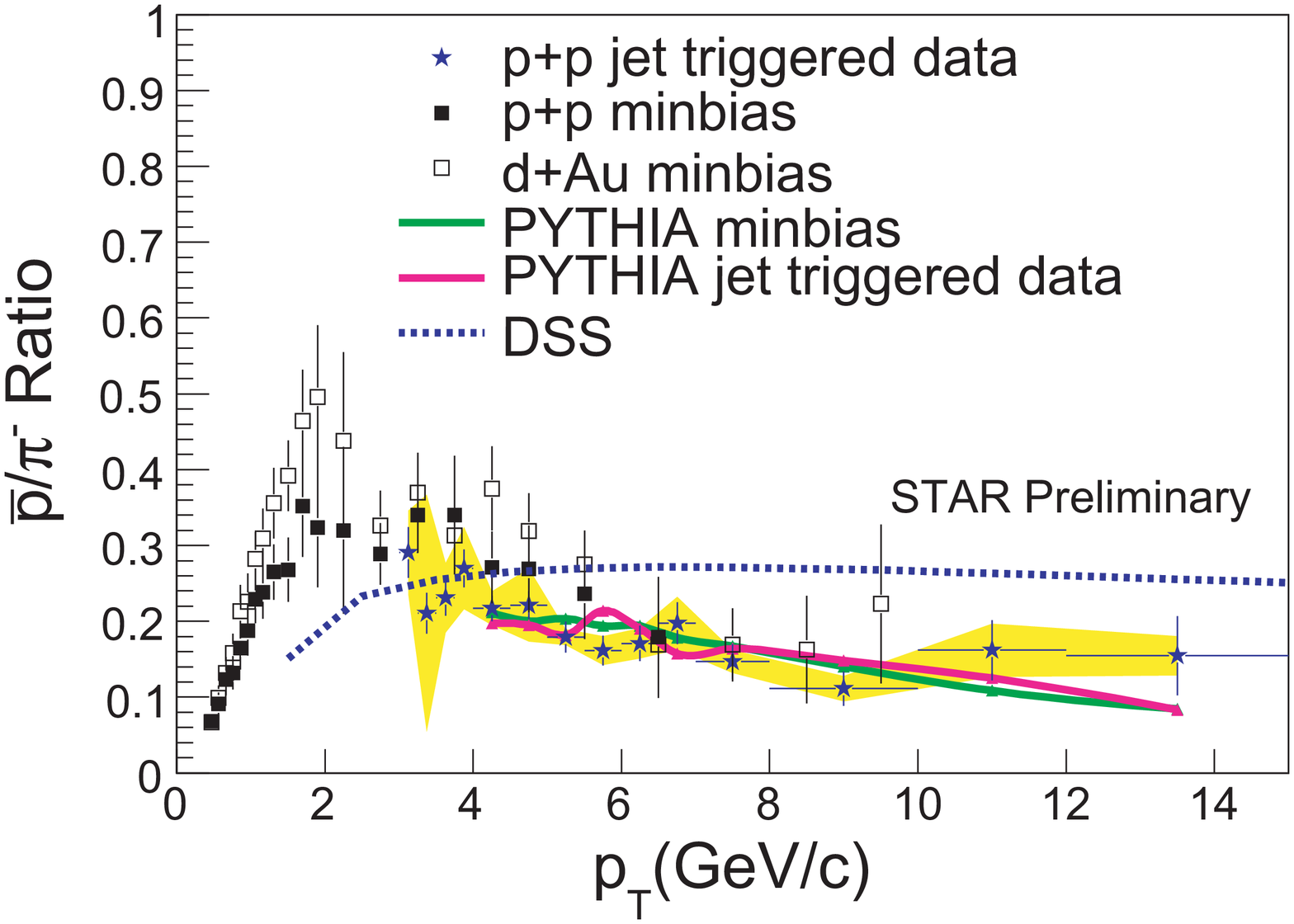}
\end{minipage}
\hspace{\fill}
\begin{minipage}[t]{85mm}
\includegraphics[height=14pc,width=18pc]{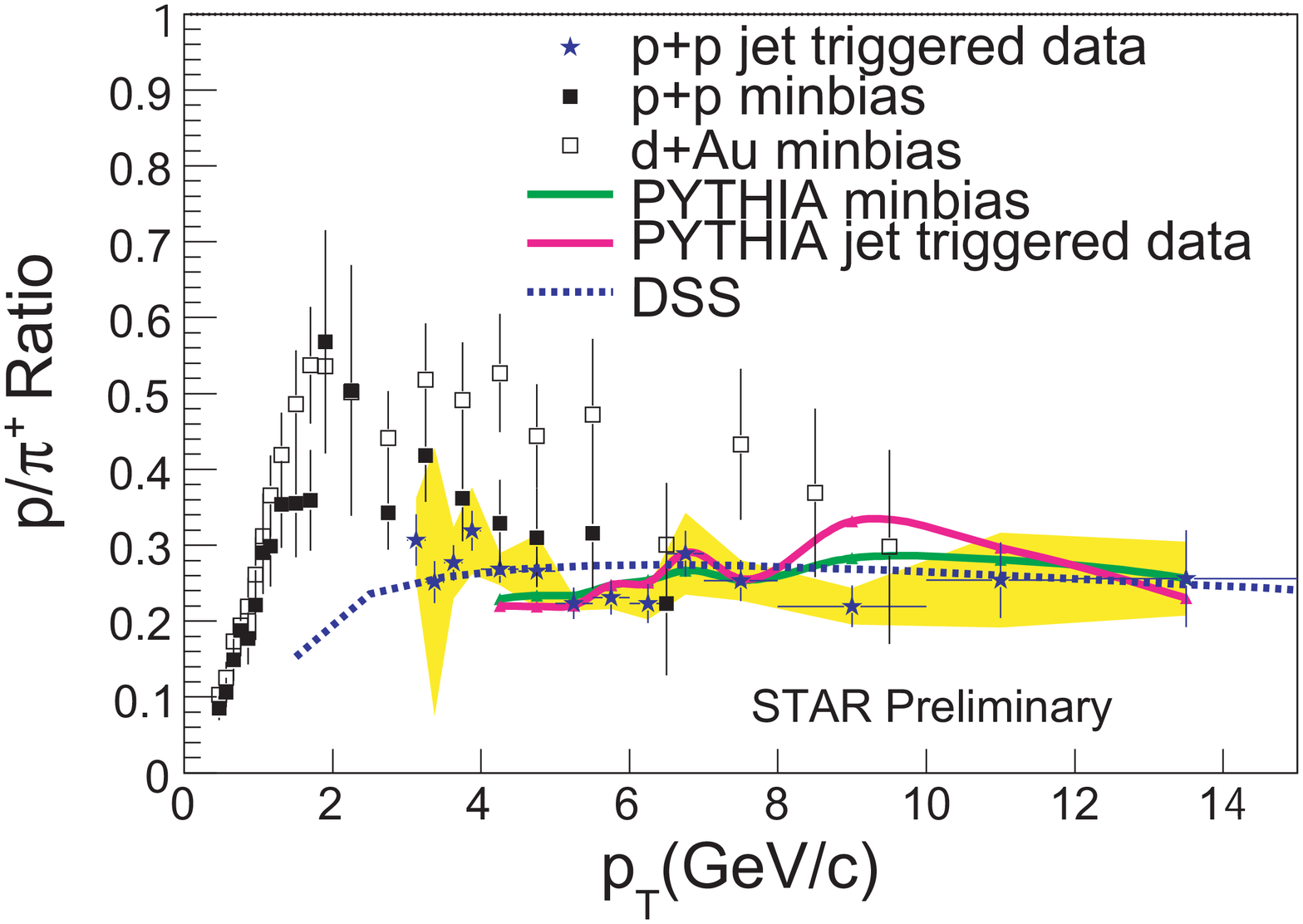}
\end{minipage}
\caption{(Color online) $\bar{p}/\pi^-$ (left panel) and $p/\pi^+$ (right panel) ratios
as a function of $p_T$ at $\sqrt{s_{_{NN}}} = 200$ GeV compared to PYTHIA \cite{pythia} and DSS \cite{dss}  NLO calculations.} \label{fig:Ratio1}
\end{figure}

\begin{figure}[htb]
\begin{minipage}[t]{85mm}
\includegraphics[height=14pc,width=18pc]{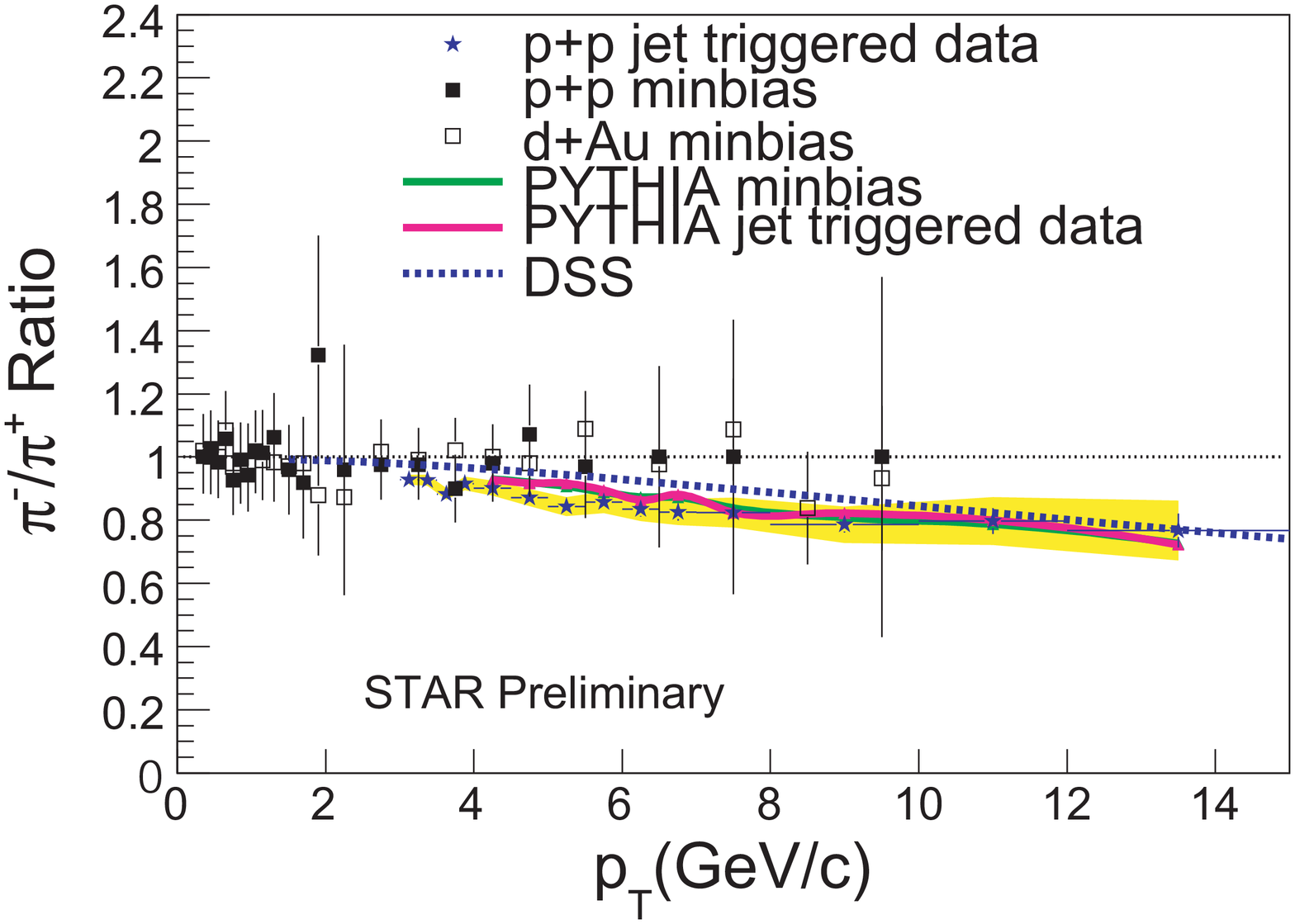}
\end{minipage}
\hspace{\fill}
\begin{minipage}[t]{85mm}
\includegraphics[height=14pc,width=18pc]{QMpbarpRatio.eps}
\end{minipage}
\caption{(Color online) $\pi^-/\pi^+$ (left panel) and $\bar{p}/p$ (right pannel) ratios
as a function of $p_T$ at $\sqrt{s_{_{NN}}} = 200$ GeV compared to PYTHIA \cite{pythia} and DSS \cite{dss} NLO calculations.} \label{fig:Ratio2}
\end{figure}

In $p+p$ collisions, charged pions and (anti-)protons were measured in the
range 5.0 $\leq p_T \leq$ 15.0 GeV/$c$ using jet triggered events. Figure \ref{fig:Ratio1}
shows the $\bar{p}/\pi^-$ (left panel) and $p/\pi^+$ (right panel) ratios
as a function of $p_T$ compared to PYTHIA \cite{pythia} (solid lines)
and DSS \cite{dss} (dotted line) NLO calculations.
In this figure, since PYTHIA minbias and PYTHIA jet triggered (two solid lines)
are the same, we can conclude that the effect of the jet trigger on the ratios is negligible.
We also observe that the DSS calculation reproduces the $p/\pi^+$ ratio; this is not the case for the $\bar{p}/\pi^-$ ratio. PYTHIA reproduces both $\bar{p}/\pi^-$ and $p/\pi^+$ ratios. Figure \ref{fig:Ratio2}
shows the $\pi^-/\pi^+$ (left panel) and $\bar{p}/p$ (right panel) ratios
as a function of $p_T$ compared to PYTHIA (solid lines) and DSS (dotted line) NLO calculations.
In this figure, we also observe that the effect of the jet trigger on the ratios is negligible and that PYTHIA reproduces both $\pi^-/\pi^+$ and $\bar{p}/p$ ratios.
Both ratios decrease as a function of $p_T$, showing the effect of valence quarks in the case of
the $\pi^-/\pi^+$ ratio and the significant quark jet contribution to the baryon production in
the case of the $\bar{p}/p$ ratio. The DSS calculation reproduces the behavior of both ratios. However, while it reproduces the magnitude of the $\pi^-/\pi^+$ ratio, it fails in the case of the $\bar{p}/p$ ratio.

\begin{figure}[htb]
\begin{minipage}[t]{85mm}
\includegraphics[height=14pc,width=18pc]{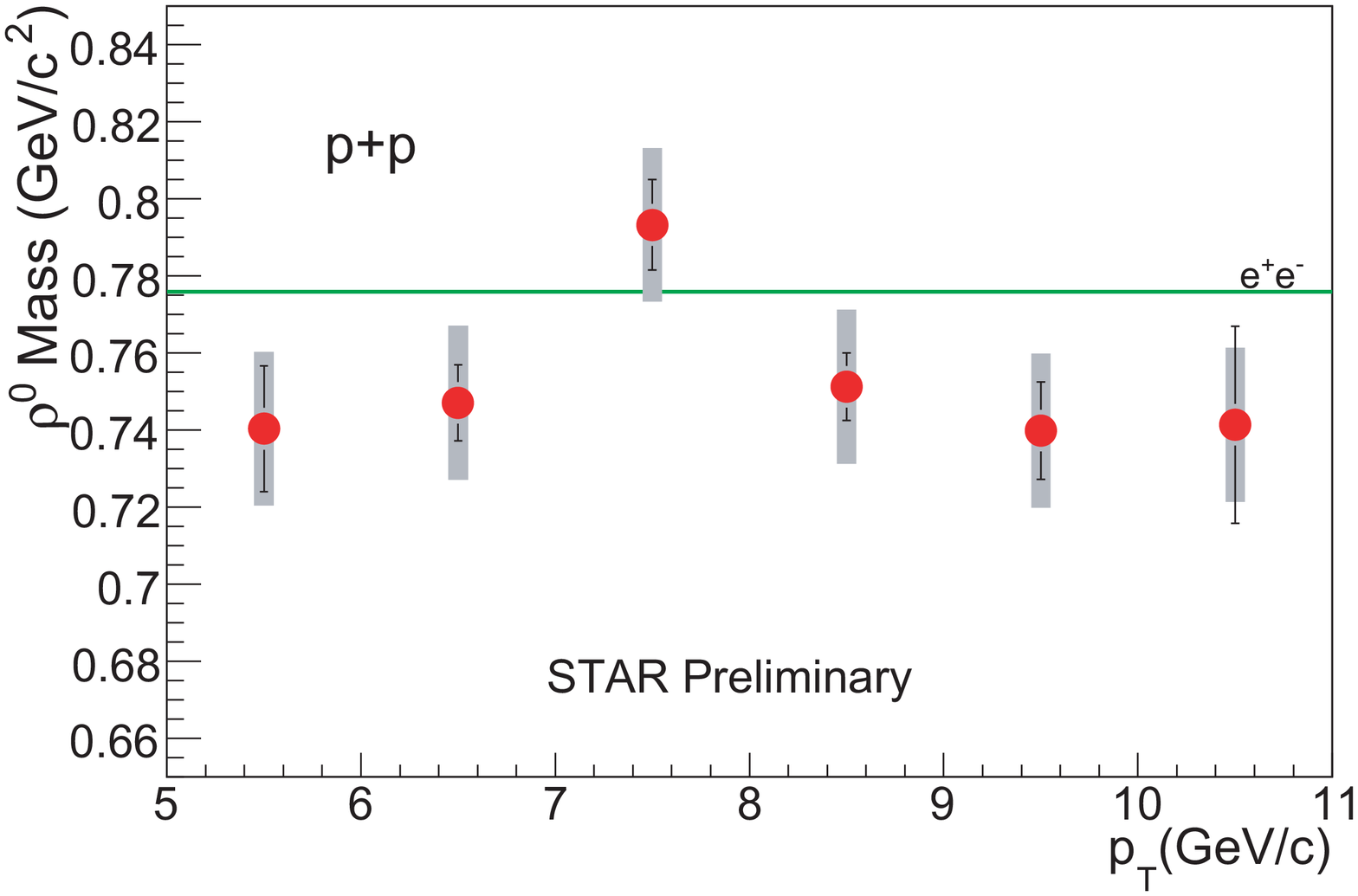}
\end{minipage}
\hspace{\fill}
\begin{minipage}[t]{85mm}
\includegraphics[height=14pc,width=18pc]{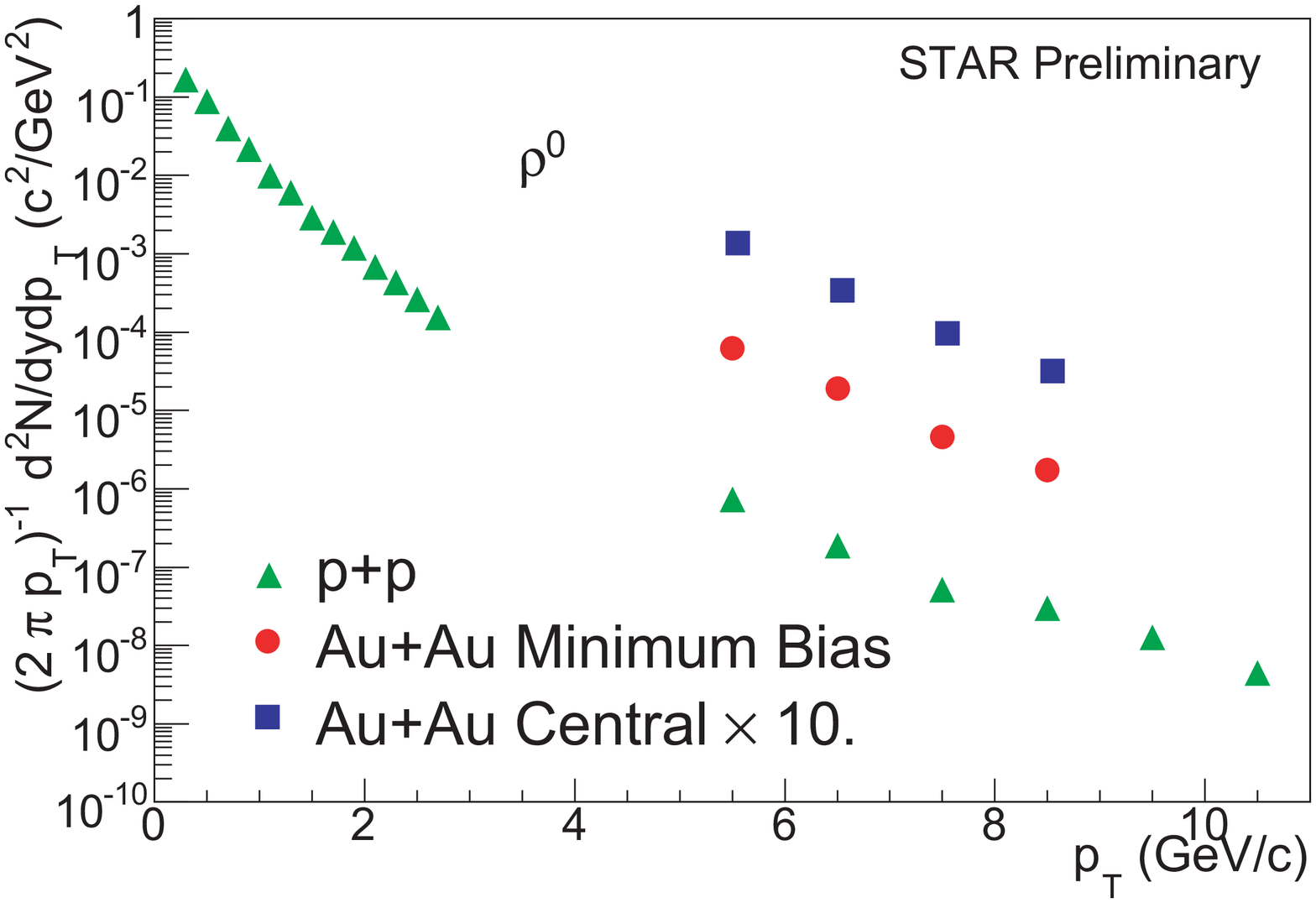}
\end{minipage}
\caption{(Color online) Left panel: $\rho^0$ mass as a function of $p_T$ measured in $p+p$ collisions at $\sqrt{s_{_{NN}}} = 200$ GeV.
The solid line corresponds to the PDG \cite{pdg} average for the $\rho^0$ measured in $e^+e^-$
interactions. Right panel: $\rho^0$ spectra as a function of $p_T$ measured in minimum bias $p+p$,
Au+Au and central Au+Au collisions at $\sqrt{s_{_{NN}}} = 200$ GeV.} \label{fig:Mass}
\end{figure}

The $\rho^0$ meson is measured at high $p_T$ (5.0 $\leq p_T
\leq$ 9.0 GeV/$c$) in minimum bias $p+p$ (using jet triggered data), Au+Au and central Au+Au collisions.
In order to obtain the transverse momentum spectra, the $\rho^0$ signal was fit to a relativistic Breit-Wigner (BW) function times the phase space
\cite{rho}. In the fits to the $\rho^0$ signal, the mass and width are fixed
according to the PDG \cite{pdg} average for the $\rho^0$ measured in $e^+e^-$
interactions, which corresponds to the rho mass in the vacuum. The mass and width of the $\rho^0$ in the vacuum reproduce well the data at high $p_T$ (5.0 $\leq p_T
\leq$ 9.0 GeV/$c$). There is not enough sensitivity to do a detailed study of the $\rho^0$ mass and width as a function of $p_T$ in minimum bias and central Au+Au collisions \cite{rho}. In the case of $p+p$ collisions, even though there is no sensitivity to do a study of the width as a function of $p_T$, we were able to look more closely into the mass. The left panel of Fig. \ref{fig:Mass} depicts the $\rho^0$
mass as a function of $p_T$ measured in $p+p$ collisions in the case the mass is a free
parameter in the fit. We observe that the $\rho^0$ mass approaches its value in the vacuum in this $p_T$ interval. The right panel of Fig. \ref{fig:Mass} shows
the $\rho^0$ transverse momentum spectra measured in minimum bias $p+p$,
Au+Au and central Au+Au collisions. The $\rho^0$ spectrum measured in $p+p$ collisions is
similar to the DSS NLO calculations for pions (within 30$\%$) as expected, since it has been shown in deep inelastic electron scattering that quarks fragment with equal probability into pions and $\rho$ mesons \cite{isr}.

\begin{figure}[htb]
\begin{minipage}[t]{85mm}
\includegraphics[height=14pc,width=18pc]{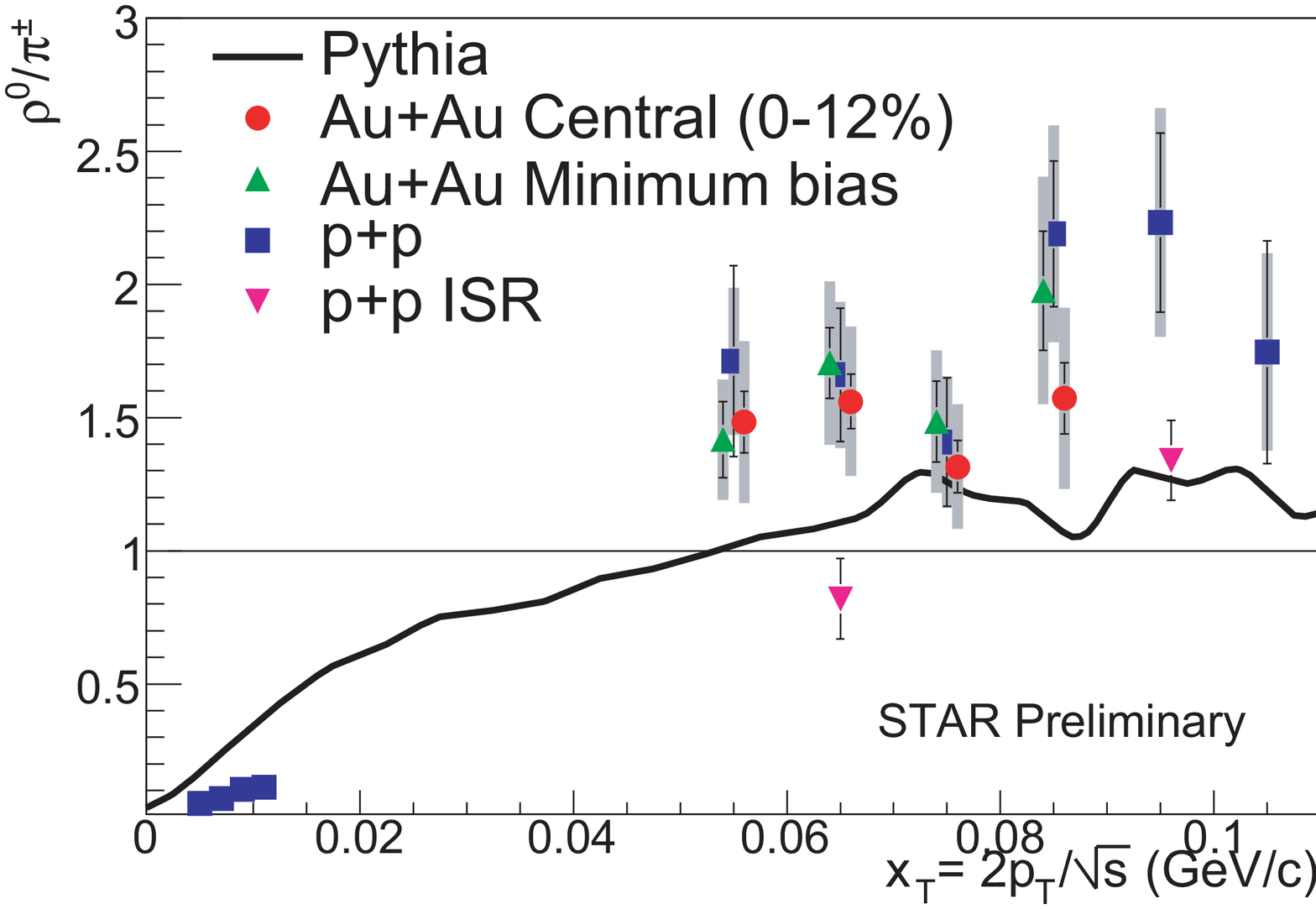}
\end{minipage}
\hspace{\fill}
\begin{minipage}[t]{85mm}
\includegraphics[height=14pc,width=18pc]{QMRhoRaa.eps}
\end{minipage}
\caption{(Color online) Left panel: The $\rho^0/\pi^-$ ratio as a function of $x_T$ measured in minimum bias $p+p$, Au+Au and
central collisions at $\sqrt{s_{_{NN}}} = 200$ GeV are compared to PYTHIA calculations. Right panel: $\pi$, $\rho^0$ and $p$ $R_{AA}$ as a function of $p_T$ at $\sqrt{s_{_{NN}}} = 200$ GeV \cite{raa}. The $\pi^0$ data comes from \cite{pi0}.} \label{fig:Raa}
\end{figure}

The $\rho^0/\pi^-$ ratio as a function of $x_T$ ($x_T = 2p_T/\sqrt{s}$) measured in minimum bias $p+p$, Au+Au and
central collisions are compared to PYTHIA calculations in the left panel of Fig. \ref{fig:Raa}.
One observes that PYTHIA under-predicts the measured ratios; however, we cannot rule out the possibility that PYTHIA can be tuned to describe the data. In the left panel of Fig. \ref{fig:Raa} it is also shown the $\rho^0/\pi^-$ measured at ISR at two different $p_T$ bins at $\sqrt{s_{_{NN}}} = 52.2$ GeV, which are lower than the STAR measurements.

The right panel of Fig. \ref{fig:Raa} depicts the charged pions, $\rho^0$ and proton plus anti-proton $R_{AA}$ \cite{raa} as a function of $p_T$. There is a separation of approximately 1.5 $\sigma$ between charged pions and protons plus anti-protons $R_{AA}$ for $p_T \geq$ 7 GeV/$c$, where the protons plus anti-protons $R_{AA}$ are larger than the charged pions $R_{AA}$. The right panel of Fig. \ref{fig:Raa} also shows that the $\pi^0$, charged pions and $\rho^0$ nuclear modification factors are comparable.

\section{Summary}

In $p+p$ collisions, charged pions
and (anti-)protons are measured in the
range 5.0 $\leq p_T \leq$ 15.0 GeV/$c$ and the $\bar{p}/\pi^+$, $p/\pi^-$,
$\bar{p}/p$ and $\pi^-/\pi^+$ ratios were discussed.
The $\rho^0$ production at high-$p_T$ (5.0 $\leq p_T \leq$
10.0 GeV/$c$) measured in minimum bias $p+p$, Au+Au and central Au+Au
collisions in the STAR detector were
presented. The $\rho^0/\pi$ ratio measured in $p+p$ is
compared to PYTHIA calculations as a test of perturbative quantum
chromodynamics (pQCD) that describes reasonably well particle
production from hard processes and also to the $\rho^0/\pi$ ratio measured in
Au+Au collisions. The charged pions, $\rho^0$ and proton plus anti-proton nuclear modification
factors were also presented. The protons plus anti-protons and the charged pions $R_{AA}$ seem to behave oppositely to what is na\"ively expected from color charge dependence of energy loss \cite{eloss}. The $\pi^0$, charged pions and $\rho^0$ nuclear modification factors are comparable indicating that the fragmentation of vector mesons and pseudo-scalars are similar in $p+p$ and Au+Au collisions.

\end{document}